\documentstyle[12pt]{article}

\newcommand{\be}{\begin{equation}}
\newcommand{\ee}{\end{equation}}
\newcommand{\ba}{\begin{eqnarray}}
\newcommand{\ea}{\end{eqnarray}}

\newcommand{\no}{\nonumber\\*[1mm]}

\newcommand{\grts}{\raise.3ex\hbox{$>$\kern-.75em\lower1ex\hbox{$\sim$}}}
\newcommand{\lets}{\raise.3ex\hbox{$<$\kern-.75em\lower1ex\hbox{$\sim$}}}

\title{\bf Solution of the strong CP problem}

\author{L.\ Lavoura \\
\small Universidade T\'ecnica de Lisboa \\
\small CFIF, Instituto Superior T\'ecnico,
Edif\'\i cio Ci\^encia (f\'\i sica) \\
\small P-1096 Lisboa Codex, Portugal}

\begin{document}

\maketitle

\begin{abstract}
I put forward an SU(2)$_L \otimes$SU(2)$_R \otimes$U(1) model
in which spontaneously broken parity symmetry
makes it that strong CP violation only arises at three-loop level.
All leptons and up-type quarks
are in doublets either of SU(2)$_L$ or of SU(2)$_R$,
but there are singlet down-type quarks.
A bi-doublet of scalars is introduced
which has only one component with non-vanishing expectation value,
thereby avoiding $W_L$-$W_R$ mixing.
The scalar potential is such that scalar-pseudoscalar mixing
does not occur either.
\end{abstract}

\section{Introduction}

Non-perturbative effects in Quantum Chromodynamics (QCD)
may lead to P and CP violation,
characterized by a parameter $\theta$,
in hadronic processes.
The experimental upper bound on the electric dipole moment of the neutron
constrains $\theta$ to be less than $10^{-9}$ or so.
The presence of this unnaturally small number in QCD
is what is known as the strong CP problem.

$\theta$ is the sum of two terms,
$\theta_{QCD}$ and $\theta_{QFD}$.
$\theta_{QCD}$ is the original value of the angle $\theta$
characterizing the QCD vacuum.
$\theta_{QFD}$ originates in the chiral rotation of the quark fields
needed to render the quark masses real and positive.
If $M_p$ and $M_n$ are the mass matrices of the up-type and down-type quarks,
then $\theta_{QFD} = \arg \det (M_p M_n)$.

There are two general ways of solving the strong CP problem.
In the first approach it is claimed that $\theta$
has no significance or physical consequences;
theories with different values of $\theta$ are equivalent,
and therefore we may set $\theta = 0$ without loss of generality.
This may be so because of the presence in the theory
of a Peccei--Quinn symmetry \cite{peccei},
but there are also claims
that QCD dynamics itself cures the strong CP problem \cite{emilio}.
The second path,
which I shall follow,
tries to find some symmetry
which naturally leads to the smallness of $\theta$.
As $\theta$ is both CP- and P-violating,
we first assume the Lagrangian
(or at least its quartic part)
to be either CP- or P-symmetric,
thereby automatically obtaining $\theta_{QCD} = 0$.
CP or P symmetry must then be either softly or spontaneously broken.
While doing this
the problem of ensuring the smallness of $\theta_{QFD}$ remains.
This is quite difficult when using CP symmetry.
After CP is broken softly or spontaneously,
it is difficult to avoid one-loop contributions to $\theta_{QFD}$
from the quark self-energies \cite{chang},
because each neutral spin-0 particle
will usually have both scalar and pseudoscalar interactions with each quark.
Georgi \cite{georgi} realized this,
and claimed that $\theta_{QFD}$ thus generated
would be of order $10^{-8}$.
However,
in order to make this estimate he assumed all quarks to have masses
at most of order 10 GeV.
Once it is known that the top mass is much larger than this,
a suppression factor is lost,
and a more likely estimate is $\theta_{QFD} \sim 10^{-5}$,
which is unacceptable.
Most models using CP to suppress $\theta$ \cite{models}
suffer from this problem in one way or another.
A remarkable exception is the model of Bento and collaborators \cite{bento}.

In 1990 the first model appeared \cite{babu}
which used P symmetry to suppress $\theta$.
The general features that models of this kind have to satisfy
were later analysed by Barr and collaborators \cite{barr}.
They claimed that a model which uses parity to suppress $\theta$
must have both mirror quarks
and a duplication of the standard-model (SM) SU(2) gauge group.
In their own words:
``The mirror families must have different weak interactions
from the ordinary $V \! - \! A$ ones.
There is simply no escape from this conclusion;
one has to double both the electroweak group and the fermionic content''.

In this paper I construct a model
in which the SU(2) gauge group is indeed doubled,
but the lepton and up-type-quark content is the same as in the SM,
while the down-type quarks are doubled.
In my model $\theta_{QFD}$ only arises at three-loop level.
The model is somehow intermediate
between the standard left-right-symmetric model \cite{review}
and the models with mirror quarks \cite{babu,barr},
but it has some odd features,
as I shall point out.

\section{The model}

\subsection{Scalar potential}

The gauge group of the model
is SU(2)$_L \otimes$SU(2)$_R \otimes$U(1).
The scalar sector consists of a bi-doublet $\phi$
(with $\tilde\phi \equiv \tau_2 \phi^\ast \tau_2$),
a doublet $\chi_L$ of SU(2)$_L$,
and a doublet $\chi_R$ of SU(2)$_R$.
The Lagrangian is assumed to be symmetric under parity,
which transforms $\phi$ into $\phi^\dagger$
and interchanges $\chi_L$ with $\chi_R$.
The gauge coupling constant of SU(2)$_L$ is equal to the one of SU(2)$_R$,
and I call it $g$.
The electric charge is $Q = T_{L3} + T_{R3} + Y$.
Up to this point,
the model is similar to the usual left-right-symmetric one \cite{review}.

I assume the Lagrangian to be symmetric under a discrete symmetry $S$,
which transforms
\be \label{Sscalar}
\phi \rightarrow i \phi,\
\chi_L \rightarrow - \chi_L,\ 
\chi_R \rightarrow i \chi_R.
\ee
As a consequence,
the scalar potential is
\ba \label{potential}
V & = &
\mu_1 (\chi_L^\dagger \chi_L + \chi_R^\dagger \chi_R)
+ \mu_2 tr (\phi^\dagger \phi)
+ m (\chi_L^\dagger \phi \chi_R + \chi_R^\dagger \phi^\dagger \chi_L)
\no
  &   &
+ \lambda_1 [(\chi_L^\dagger \chi_L)^2 + (\chi_R^\dagger \chi_R)^2]
+ \lambda_2 (\chi_L^\dagger \chi_L) (\chi_R^\dagger \chi_R)
+ \lambda_3 [tr (\phi^\dagger \phi)]^2
\no
  &   &
+ \lambda_4 tr (\phi^\dagger \tilde\phi) tr (\tilde\phi^\dagger \phi)
+ \lambda_5 \{ [tr (\phi^\dagger \tilde\phi)]^2
+ [tr (\tilde\phi^\dagger \phi)]^2 \}
\no
  &   &
+ \lambda_6 (\chi_L^\dagger \chi_L + \chi_R^\dagger \chi_R)
tr (\phi^\dagger \phi)
\no
  &   &
+ \lambda_7 (\chi_L^\dagger \phi \phi^\dagger \chi_L
+ \chi_R^\dagger \phi^\dagger \phi \chi_R)
+ \lambda_8 (\chi_L^\dagger \tilde\phi \tilde\phi^\dagger \chi_L
+ \chi_R^\dagger \tilde\phi^\dagger \tilde\phi \chi_R).
\ea
It is important to notice that,
even though I do not impose CP symmetry,
all the coupling constants in this $S$- and parity-symmetric potential
are real.
I decompose the scalar multiplets as ($A$ may be $L$ or $R$)
\be \label{scalardecomposition}
\phi =
\left(\!\!
\begin{array}{cc}
\phi_2^{0 \ast} & \phi_1^+ \\
- \phi_2^- & \phi_1^0
\end{array}
\!\right),\ \
\tilde\phi =
\left(\!\!
\begin{array}{cc}
\phi_1^{0 \ast} & \phi_2^+ \\
- \phi_1^- & \phi_2^0
\end{array}
\!\right),\ \
\chi_A =
\left(\!
\begin{array}{c}
\chi_A^+ \\
\chi_A^0
\end{array}
\!\right).
\ee
In principle,
all four neutral complex fields in these multiplets may acquire
a vacuum expectation value (vev).
I shall however assume the vev of $\phi_2^0$ to vanish.
That is,
I assume
\be \label{vevs}
\langle \phi \rangle =
\left(\!
\begin{array}{cc}
0 & 0 \\
0 & k_1
\end{array}
\!\right),\ \
\langle \tilde\phi \rangle =
\left(\!
\begin{array}{cc}
k_1^\ast & 0 \\
0 & 0
\end{array}
\!\right),\ \
\langle \chi_A \rangle =
\left(\!
\begin{array}{c}
0 \\
v_A
\end{array}
\!\right).
\ee
This is a crucial feature of the model.
The purpose of this is to avoid $W_L$-$W_R$ mixing which,
together with the different mixing matrices
for left- and right-handed currents
if $\langle \phi_2^0 \rangle$ were non-vanishing,
would lead to the generation of $\theta_{QFD}$ at one-loop level.
It is important to notice that $\langle \phi_2^0 \rangle = 0$
is a consistent assumption
since no vev for $\phi_2^0$ is induced by the vevs of $\phi_1^0$,
$\chi_L^0$ and $\chi_R^0$.
This means that,
when we substitute all scalar fields but $\phi_2^0$ by their vevs,
no term remains in the potential which is linear in $\phi_2^0$.
I must add the remark that the potential in eq.~\ref{potential}
is the most general one which has parity symmetry
and admits $\langle \phi_2^0 \rangle = 0$.

Because of the gauge symmetry,
the only meaningful phase in eq.~\ref{vevs}
is $\alpha \equiv \arg (v_L v_R^\ast k_1^\ast)$.
The value of $\alpha$ is fixed by the only term in the potential
in eq.~\ref{potential} which ``sees'' it,
the term in $m$.
However,
$m$ is real,
and therefore
(assuming,
without loss of generality,
$m$ to be negative)
$\alpha$ will be zero.

As there is only one term in the potential
which ``sees'' the vacuum phase $\alpha$,
there is no scalar-pseudoscalar mixing.
This is another crucial feature of the model.
It prevents one-loop self-energy diagrams with neutral spin-0 fields
from generating $\theta_{QFD}$,
which would then be too large.

The absence of scalar-pseudoscalar mixing can be explicitly checked
by developing the potential in eq.~\ref{potential}
to find the masses of the various spin-0 particles.
Without loss of generality I set $k_1$,
$v_L$ and $v_R$ real and positive.
I then write
\be \label{neutraldecomposition}
\phi_1^0 = k_1 + \frac{\rho_1 + i \eta_1}{\sqrt{2}},\
\phi_2^0 = \frac{\rho_2 + i \eta_2}{\sqrt{2}},\
\chi_A^0 = v_A + \frac{\rho_A + i \eta_A}{\sqrt{2}}.
\ee
There is one massive neutral pseudoscalar,
$ I \equiv (k_1 v_R \eta_L - k_1 v_L \eta_R - v_L v_R \eta_1) / N $,
where $N \equiv \sqrt{k_1^2 v_R^2 + k_1^2 v_L^2 + v_L^2 v_R^2}$
is a normalization factor.
The mass of $I$ is proportional to $m$,
as would be expected,
because if $m$ vanished
there would be another spontaneously broken U(1) symmetry
(given by $\chi_L \rightarrow \exp (i \psi) \chi_L$
and $\chi_R \rightarrow \exp (-i \psi) \chi_R$)
in the potential,
and then $I$ would be a Goldstone boson.
There are three massive neutral scalars,
which are orthogonal combinations of $\rho_1$,
$\rho_L$ and $\rho_R$.
These orthogonal combinations depend on the specific values
of the parameters in the potential.
There are two other neutral spin-0 particles,
$\rho_2$ and $\eta_2$.
The masses of $\rho_2$ and $\eta_2$
are different because of the $\lambda_5$ term in the potential.
Finally,
the physical charged scalars are
$H_L^+ = (- k_1 \chi_L^+ + v_L \phi_1^+) / V_L$ and
$H_R^+ = (- k_1 \chi_R^+ - v_R \phi_2^+) / V_R$,
where $V_A \equiv \sqrt{k_1^2 + v_A^2}$
are real and positive normalization factors.
$H_L$ and $H_R$ do not mix,
and this is another distinguishing feature of the $S$-symmetric potential
in eq.~\ref{potential}.

\subsection{Gauge bosons}

Each SU(2) gauge group gives rise to a charged gauge boson.
Because $\langle \phi_2^0 \rangle$ vanishes,
the two charged gauge bosons do not mix.
The mass of $W_L$ is $ (g V_L) / \sqrt{2} $
and the mass of $W_R$ is $ (g V_R) / \sqrt{2} $.

As usual,
$W_L$ is identified with the observed $W$ particle.
In order to have the mass of the $W_R$ much larger than the one of the $W_L$,
we must impose the condition $v_R \gg v_L, k_1$.
This condition also ensures \cite{review}
that the two massive neutral gauge bosons are such that
the lightest of them has a mass which is approximately
$ (g V_L) / (\sqrt{2} \cos \theta_W) $,
and its interactions are approximately given by
$ (g /\!\cos \theta_W) (T_{L3} - Q \sin^2 \theta_W) $,
with a suitably defined angle $\theta_W$.
The above approximations
are up to terms of order $v_L / v_R$ or $k_1 / v_R$.
This means that the lightest neutral gauge boson can be identified
with the $Z$ particle observed at LEP and SLAC if $v_R \gg v_L, k_1$.
Notice however that we do not impose any condition on $k_1 / v_L$.

\subsection{Yukawa interactions}

The quark sector of the model consists of doublets $q_L$ of SU(2)$_L$,
doublets $q_R$ of SU(2)$_R$,
and left-handed $N_L$ and right-handed $n_R$
which are singlets both of SU(2)$_L$ and of SU(2)$_R$
and have electric charge $-1/3$.
There are three (for three generations)
of each of these types of quark multiplets.
I write
\be \label{quarkdecomposition}
q_L =
\left(\!
\begin{array}{c}
p_L \\
n_L
\end{array}
\!\right),\
q_R =
\left(\!
\begin{array}{c}
p_R \\
N_R
\end{array}
\!\right).
\ee
Parity interchanges $q_L$ with $q_R$ and $N_L$ with $n_R$. 
The discrete symmetry $S$ transforms
\be
q_L \rightarrow q_L,\
q_R \rightarrow i q_R,\
N_L \rightarrow N_L,\
n_R \rightarrow - n_R.
\ee
As a consequence,
the Yukawa Lagrangian for the quarks is
\ba \label{Yukawa}
\cal{L}_Y & = &
- \overline{q_L} \tilde\phi \Delta q_R
- \overline{q_R} \tilde\phi^\dagger \Delta q_L
\no
          &   &
- (\overline{q_L} \chi_L G n_R + \overline{q_R} \chi_R G N_L + H.c.),
\ea
where $\Delta$ is hermitian,
while $G$ is a general 3$\times$3 complex matrix.
The mass terms will then be
\be \label{masses}
\cal{L}_Y = ... - (\overline{p_L} k_1^\ast \Delta p_R
+ \overline{n_L} v_L G n_R
+ \overline{N_L} v_R^\ast G^\dagger N_R
+ H.c.).
\ee
It is clear that at tree-level
$\theta_{QFD} = \arg \det [(k_1^\ast \Delta)(v_L G)(v_R^\ast G^\dagger)] =
\alpha + \arg \det \Delta + \arg \det (G G^\dagger) = 0$,
i.e.,
strong CP violation vanishes at tree level.
It is important to stress that in this model,
contrary to what happened in most previous ones,
the value of the vacuum phase ($\alpha = 0$) is crucial
to obtain $\theta_{QFD} = 0$ at tree level.
In particular,
in the previous models using parity symmetry \cite{babu,barr}
there is no gauge-invariant vacuum phase at all,
just as happens in the SM.
In the present model the scalar potential is very important,
because $\alpha = 0$ and $\theta_{QFD} = 0$ rely on it.

Without loss of generality,
we may set the vevs to be real and positive as before,
and choose a basis for the doublets $q_L$ and $q_R$
in which $\Delta = M_u / k_1$ is a real,
diagonal and positive matrix.
$M_u = {\rm diag} (m_u, m_c, m_t)$
is the diagonal matrix of the masses of the up-type quarks.
In this basis,
$p_L = u_L$ and $p_R = u_R$ are the up-type-quark fields.
We bi-diagonalize the matrix $G$ to obtain
$M_d = {\rm diag} (m_d, m_s, m_b)$,
the diagonal matrix of the masses of the down-type quarks:
$V^\dagger G U = M_d / v_L$.
$V$ is the Cabibbo--Kobayashi--Maskawa (CKM) matrix.
The CKM matrix is identical for both the left-handed
and the right-handed charged gauge interactions.
Indeed,
with
\be \label{quarkrotations}
n_L = V d_L,\
n_R = U d_R,\
N_L = U D_L,\
N_R = V D_R,
\ee
those interactions are given by the following terms in the Lagrangian:
\be \label{gaugeinteraction}
\frac{g}{\sqrt{2}}\,
( W_{L \mu}^+ \overline{u_L} \gamma^\mu \gamma_L V d_L
+ W_{R \mu}^+ \overline{u_R} \gamma^\mu \gamma_R V D_R
+ H.c.).
\ee
$\gamma_R = (1 + \gamma_5) / 2$
and $\gamma_L = (1 - \gamma_5) / 2$
are the chiral projection matrices.
The usual down-type quarks,
with mass matrix $M_d$,
have chiral components $d_L$ and $d_R$,
and there are three exotic down-type quarks,
with components $D_L$ and $D_R$,
the mass matrix of which is $(v_R / v_L) M_d$.
This means that the masses of the $D$ quarks
are proportional to the masses of the normal down-type quarks,
the proportionality factor being $v_R/v_L \gg 1$.

After diagonalizing the quark mass matrices,
the Yukawa interactions of the quarks
with the physical charged spin-0 fields are given by
\ba \label{chargedyukawas}
\cal{L}_Y & = & ...
+ \frac{H_L^+}{V_L}\, \overline{u}
\left(
\frac{v_L}{k_1} M_u V \gamma_L + \frac{k_1}{v_L} V M_d \gamma_R
\right)
d
\no
          &   &
+ \frac{H_R^+}{V_R}\, \overline{u}
\left(
\frac{v_R}{k_1} M_u V \gamma_R + \frac{k_1}{v_L} V M_d \gamma_L
\right)
D
+ H.c.,
\ea
while the Yukawa interactions of the quarks with $\rho_2$ and $\eta_2$
are given by
\ba \label{rho2yukawas}
\cal{L}_Y & = & ...
- \frac{\rho_2}{\sqrt{2} k_1}
\left(
\overline{D} V^\dagger M_u V \gamma_L d
+ \overline{d} V^\dagger M_u V \gamma_R D
\right)
\no
          &   &
+ \frac{i \eta_2}{\sqrt{2} k_1}
\left(
\overline{D} V^\dagger M_u V \gamma_L d
- \overline{d} V^\dagger M_u V \gamma_R D
\right).
\ea
The Yukawa interactions of the quarks with $I$
and with the scalars are given by
\ba \label{scalaryukawas}
\cal{L}_Y & = & ...
- \frac{i I}{\sqrt{2} N}
\left(
\frac{v_L v_R}{k_1} \overline{u} M_u \gamma_5 u +
\frac{k_1 v_R}{v_L} \overline{d} M_d \gamma_5 d +
k_1 \overline{D} M_d \gamma_5 D
\right)
\no
          &   &
- \frac{\rho_1}{\sqrt{2} k_1} \overline{u} M_u u
- \frac{\rho_L}{\sqrt{2} v_L} \overline{d} M_d d
- \frac{\rho_R}{\sqrt{2} v_L} \overline{D} M_d D.
\ea
Remember that $\rho_1$,
$\rho_L$ and $\rho_R$ are not eigenstates of mass,
rather they are related by an orthogonal transformation
to the three physical scalar fields.

The lepton sector of the model
may be chosen to be the usual one \cite{review},
without lepton singlets.
This is a considerable simplification relative to previous models
which used parity to suppress $\theta_{QFD}$ \cite{babu,barr}.
Because the vev of $\phi_2^0$ vanishes,
the neutrino masses vanish too.
In this way we do not have to explain their small value,
which is a usually a problem in left-right-symmetric models.

\subsection{CP violation}

In this model CP violation is hard
and manifests itself only in the complexity of the CKM matrix $V$.
At lowest order in $V$,
rephasing-invariant imaginary parts appear in the ``quartets''
$V_{\alpha i} V_{\beta j} V_{\alpha j}^\ast V_{\beta i}^\ast$
($\alpha \neq \beta$ and $i \neq j$).
As is well-known \cite{V},
the imaginary parts of all quartets are equal in modulus to a value $J$.
The experimental data on $V$ imply that $J$ is at most $10^{-4}$.

Let us consider the generation of $\theta_{QFD}$ from the quark self-energies.
Just as in the SM \cite{ellis},
complex self-energies arise at lowest order
when they are proportional to quartets.
This happens when the respective diagrams
have four vertices with either the gauge bosons $W$
or the charged spin-0 particles $H_L$ or $H_R$.
This means that complex self-energies only arise at two-loop level.

In the unitary gauge,
the self-energies with gauge bosons $W$ do not contribute to the masses
--- they are proportional to either $\not \! p \gamma_L$
or $\not \! p \gamma_R$.
In other gauges though,
the Goldstone bosons absorbed in the longitudinal components of the $W$
contribute an imaginary part to the mass of each quark.
However,
when divided by the mass of the quark and summed over all flavours,
those imaginary parts cancel out.
As a consequence the contribution to $\theta_{QFD}$ vanishes in any gauge.
Using a similar argument,
we can easily show that the two-loop diagrams involving the spin-0
charged fields $H_L$ and $H_R$ do not contribute to $\theta_{QFD}$ either.

Therefore,
strong CP violation only arises at three-loop level.
This is the same as happens in the SM \cite{ellis}.
$\theta$ will be suppressed by the factors
$J \lets 10^{-4}$ and $(16 \pi^2)^{-3} \sim 10^{-6}$.
Therefore,
we expect $\theta$ to be no larger than $10^{-10}$.
However,
as was pointed out in the analysis of Ellis and Gaillard \cite{ellis},
factors $m_q / m_W$ should be present too,
where $m_q$ are second-generation quark mases.
These factors $m_q/m_W$ further suppress $\theta_{QFD}$
to values of order $10^{-16}$.
Moreover,
in models which use parity symmetry to obtain $\theta_{QFD} = 0$ at tree level,
there must be extra suppression factors at loop level
arising from a partial cancellation
between the contributions to $\theta_{QFD}$
from the usual quarks and from the mirror sector \cite{barr}.
Therefore,
$10^{-16}$ is only an upper bound for $\theta_{QFD}$,
which in the present model is probably much smaller than that.

\section{Conclusions}

It is important to stress the differences between the present model
and previous ones \cite{babu,barr}
which used parity symmetry to suppress $\theta_{QFD}$.

This model has less fermions,
because it avoids doubling the lepton spectrum
and the up-type-quark spectrum.
On the other hand,
it has more spin-0 fields,
because of the use of a bi-doublet of SU(2)$_L$ and SU(2)$_R$.
The presence of a discrete symmetry is crucial to obtain
$\langle \phi_2^0 \rangle = 0$;
this in turn is important
because it allows us to have vanishing neutrino masses.
The asymmetry between up-type and down-type quarks,
with twice as many charge $-1/3$ quarks as charge $2/3$ quarks,
is an interesting feature of the model;
of course,
this asymmetry may be inverted,
we might instead consider a model in which the number of up-type quarks
would be twice the number of down-type quarks.

In the previous models,
the ratio between the masses of the $W_R$ and of the $W_L$ was $v_R / v_L$,
which was equal to the ratio
between the masses of the heavy and light down-type quarks,
and of the heavy and light up-type quarks.
In the present model,
$v_R / v_L$ is still the ratio between the masses
of the heavy and light down-type quarks.
On the other hand,
the ratio of the masses of $W_R$ and $W_L$ is now
$V_R / V_L \neq v_R / v_L$.

The ratio $v_R / v_L$ must at least be of order $10^5$,
else the lightest exotic $D$ quark would have been observed at LEP.
However,
the mass of $W_R$ does not have to be
that much higher than the one of the $W_L$,
because $k_1 / v_L$ may be quite large,
and then $V_R / V_L \sim v_R / k_1$ instead of $v_R / v_L$.
It is worth noting that the mass matrix of the charge $2/3$ quarks
is proportional to $k_1$,
while the one of the charge $-1/3$ quarks is proportional to $v_L$
(see eq.~\ref{masses});
this fact suggests that $k_1 / v_L \sim 100$ may be a good guess.

The discrete symmetry is needed,
not only to make $\langle \phi_2^0 \rangle = 0$ consistent,
but also in order to make all couplings in the scalar potential real,
even when CP symmetry is not imposed to the Lagrangian,
and to guarantee that no scalar-pseudoscalar mixing will arise.
That symmetry is also necessary to lead to a vanishing vacuum phase $\alpha$.
Contrary to most previous models,
the fact that the vacuum phase vanishes
is crucial to obtain $\theta_{QFD} = 0$ at tree level.

This model seems to be a viable and interesting alternative
to the usual left-right-symmetric model,
attractive in particular
because of the absence of flavour-changing neutral interactions
(except for the ones mediated by $\rho_2$ and by $\eta_2$,
which connect the standard down-type quarks with the heavy ones,
see eq.~\ref{rho2yukawas}).
Its experimental exploration,
searching in particular
for the right-handed gauge interactions of the up-type quarks
and for the effects of the charged spin-0 fields,
should be given some attention.


\begin{thebibliography}{99}

\bibitem{peccei}
R.\ Peccei and H.\ Quinn,
Phys.\ Rev.\ Lett.\ 38 (1977) 1440
and
Phys.\ Rev.\ D 16 (1977) 1791.

\bibitem{emilio}
P.\ Bicudo and J.\ Ribeiro,
preprint FISIST/2-96/CFIF (hep-ph/9604219).

\bibitem{chang}
D.\ Chang and R.\ N.\ Mohapatra,
Phys.\ Rev.\ D 32 (1985) 293.

\bibitem{georgi}
H.\ Georgi,
Hadr.\ J.\ 1 (1978) 155.

\bibitem{models}
R.\ N.\ Mohapatra and G.\ Senjanovi\'c,
Phys.\ Lett.\ 79B (1978) 283;
M.\ A.\ B.\ B\'eg and H.-S.\ Tsao,
Phys.\ Rev.\ Lett.\ 41 (1978) 278;
G.\ Segr\`e and H.\ A.\ Weldon,
Phys.\ Rev.\ Lett.\ 42 (1979) 1191;
S.\ Barr and P.\ Langacker,
Phys.\ Rev.\ Lett.\ 42 (1979) 1654;
G.\ C.\ Branco, W.\ Grimus and L.\ Lavoura,
Phys.\ Lett.\ B 380 (1996) 119.

\bibitem{bento}
L.\ Bento, G.\ C.\ Branco and P.\ Parada,
Phys.\ Lett.\ B 267 (1991) 95.

\bibitem{babu}
K.\ S.\ Babu and R.\ N.\ Mohapatra,
Phys.\ Rev.\ D 15 (1990) 1958.

\bibitem{barr}
S.\ M.\ Barr, D.\ Chang and G.\ Senjanovi\'c,
Phys.\ Rev.\ Lett.\ 67 (1991) 2765. 

\bibitem{review}
For a review see G.\ Senjanovi\'c,
Nucl.\ Phys.\ B 153 (1979) 334.

\bibitem{V}
C.\ Jarlskog,
Phys.\ Rev.\ Lett.\ 55 (1985) 1039
and Z.\ Phys.\ C 29 (1985) 491;
G.\ C.\ Branco and L.\ Lavoura,
Phys.\ Lett.\ B 208 (1988) 123.

\bibitem{ellis}
J.\ Ellis and M.\ K.\ Gaillard,
Nucl.\ Phys.\ B 150 (1979) 141.

\end{thebibliography}
\end{document}